\documentclass[12pt,preprint]{aastex}

\newcommand{\eg}{{\it e.g.\ }}
\newcommand{\etal}{{ et al.~\/}}
\newcommand{\ie}{{\it i.e.~\/}}
\newcommand{\mf}{{Minkowski functionals }}
\newcommand{\qmask}{{QMASK }} 
\newcommand{\dt}{{\Delta T }}

\begin{document}

\title{Morphological Measures of nonGaussianity in CMB Maps}
\author{Sergei F. Shandarin and Hume A. Feldman}
\affil{Department of Physics and Astronomy, \ University of Kansas, 
\ Lawrence, KS 66045}
 
\author{Yongzhong Xu, and Max Tegmark}
\affil{Department of Physics and Astronomy, \ University of Pennsylvania,
\ Philadelphia, PA 19104}

\begin{abstract}
We discuss the tests of nonGaussianity in CMB maps using morphological
statistics known as Minkowski functionals.
As an example we test degree-scale 
cosmic microwave background (CMB) anisotropy for Gaussianity by 
studying the \qmask map that was obtained from combining
the QMAP and Saskatoon data. 
We compute seven morphological functions $M_i(\dt)$, $i=1,...,7$: 
six \mf and the number of regions $N_c$ at a hundred $\dt$
levels. We also introduce a new
parameterization of the morphological functions $M_i(A)$  
in terms of the total area $A$ of the excursion set. 
We show that the latter considerably
decorrelates the morphological statistics 
and makes them more robust because they are less sensitive
to the measurements at extreme levels.
We compare these results with those from 1000 Gaussian 
Monte Carlo maps with the same sky coverage, noise properties 
and power spectrum, and
conclude that the \qmask map is neither a very typical nor a very exceptional
realization of a Gaussian field. At least about 20\% of the 1000
Gaussian Monte Carlo maps differ more than the \qmask map from the mean 
morphological parameters of the Gaussian fields. 
\end{abstract}

\section{Introduction}
The issue of Gaussianity of CMB maps plays a crucial role in
testing assumptions about the early Universe. The simplest inflation
models strongly favor Gaussianity of the primordial inhomogeneities
\citep[see][and references therein]{tur97}, whereas other scenarios 
assuming cosmic strings or topological defects 
\citep[see][and references therein]{vil-she94} 
predict non-Gaussian perturbations. 
Gaussianity is also a key underlying assumption of all experimental 
power spectrum analyses to date, entering into the computation of 
error bars \citep{teg97,bon-jaf98}, and therefore needs to be 
observationally tested.
A third reason for studying Gaussianity of CMB maps 
is that it may reveal otherwise undetected foreground contamination.

Numerous tests of Gaussianity in the COBE maps 
\citep{col-etal96,kog-etal96,fer-mag-gor98,pan-val-fan98,bro-teg99,nov-fel-sh99,ban-zar-gor00,bar-etal00,mag00,muk-hob-las00,agh-for-bou01,phi-kog01} 
have resulted in the general agreement that all non-Gaussian signals
were of noncosmological origin. This was not unexpected because of COBE's
low ($7^\circ$ deg) angular resolution.

Testing Gaussianity on smaller scales may bring much more 
interesting results. The first study of Gaussianity on degree
scale showed the consistency of the \qmask map with the assumption
of the Gaussianity \citep{par-etal01}. However, this study tested
only the Gaussianity of the topology of the map.
The study of the MAXIMA-1 anisotropy data also showed the consistency
with the Gaussianity in the range between 10 arc-minutes and 5 degrees
\citep{wu-etal01}.
A total of 82 tests for Gaussianity were made in this study.
Although performing many tests is always better than few it is not 
clear how independent these tests were. 

The question of 
statistical independence of different non-Gaussianity tests 
is complicated. However, in some simple cases some of the tests
can be statistically disentangled.
For instance, consider the probability function, which is  
perhaps the simplest test for non-Gaussianity.
It easy to transform the probability
function to any given form simply by relabeling the contours using
a monotonic function. This relabeling obviously does not have any effect
on the morphology of the field since every contour line remains the same
as well as the order of levels due to monotonicity of the transformation.
However, this transformation may strongly affect the whole hierarchy 
of the $n$-point correlation functions. This is of course due to 
well known connection
of the probability function to all $n$-point functions.

Genus represents only one statistic sensitive to non-Gaussianity.
It may detect some types of non-Gaussianity and be not sensitive to
the others. Consider, for instance, the following mapping of a
Gaussian field $\phi(x,y)$  into  $\Phi(X,Y)$ 
\begin{equation}
\Phi(X,Y) = \phi(x,y),
\end{equation}
where
\begin{eqnarray}
X&=&x+\tau u(x,y), \nonumber \\
Y&=&y+\tau v(x,y). \label{mapping}
\end{eqnarray}
Equation 1 guarantees that every level label remains unchanged,
while the level lines are shifted and deformed according to eqs.~\ref{mapping}.
Functions $u(x,y)$ and $v(x,y)$ can be any smooth functions including
random functions. At small $\tau$ before caustics 
have occurred, the mapping $(x,y) \rightarrow (X,Y)$ is single valued and
therefore the contour map of  $\Phi(X,Y)$ is one-to-one image of $\phi(x,y)$,
although geometrically distorted. The genus of the $\Phi(X,Y)$ field as
a function of the level remains exactly the same (\ie Gaussian) as that
of the $\phi(x,y)$ field because the mapping is continous and nonsingular
and therfore preserves toplogy. On the other hand, the mapping 
$(x,y) \rightarrow (X,Y)$ can be neither isometric (conserving the lengths)
nor area-preserving and therefore both the areas and the contours of the
excursion sets may be distorted. As a result, the  $\Phi(X,Y)$ can be
a strongly non-Gausssian field having exactly the Gaussian genus
in the level parameterization.  Perhaps, it is worth noting 
that a particular case of the above example 
($u(x,y)=\partial \psi(x,y)/\partial x$ and 
$v(x,y)=\partial \psi(x,y)/\partial y$)
corresponds  to viewing 
the Gaussin contour map $\phi(x,y)$  through a glass layer with
varying thickness $\psi(x,y)$ (see \eg \cite{z-mam-sh83,sh-z89}).

\begin{figure}
\includegraphics[height=15cm]{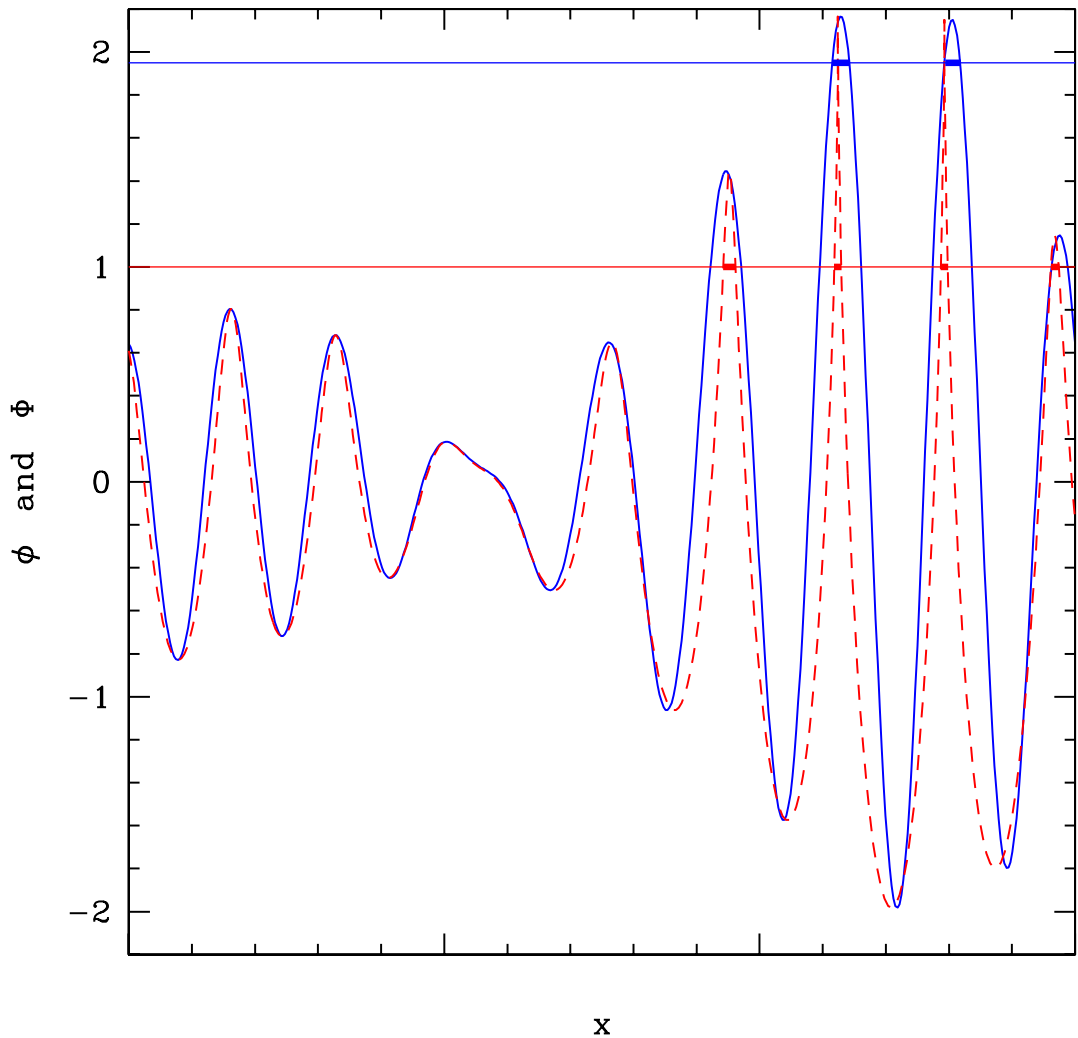}
\caption{One-dimensional illustration of a non-Gaussian field $\Phi$
(dashed line) having exactly same number of peaks (which is analogous
to genus in 2D) in the level parameterization as the Gaussian field
$\phi$ (solid line). Compare the number of
peaks in two fields at two levels marked by the dotted and dash-dotted 
horizontal lines. The length parameterization in 1D is analogous to
the area parameterization in 2D. In order to compare the number of
peaks in the length parameterization one has to count the peaks of
the two fields at different levels. The marked levels are chosen in such 
a way that the total length of the excursion set of the non-Gaussian field
at $\Phi=1.$ (the sum of four heavy segments) equals the total length of the 
the excursion set of the Gaussian field at $\phi=1.95$ 
(the sum of two heavy segments).}  \label{OneDGenus}
\end{figure}

A one-dimensional illustration of the above field is shown in 
Fig. \ref{OneDGenus}.
A realization of the Gaussian field $\phi(x)$ is shown by the solid
line and of the non-Gaussian field $\Phi(x)$ by the dashed line.
The total area of the excursion set in 2D is analogous to the total
length of the excursion set in 1D and the genus in 2D is analogous
to the number of peaks in 1D. The level parameterization means the
comparison of the number of peaks in the two fields at the same level.
Both fields have the same number of peaks at every level
as illustrated by the dotted and dash-dotted horizontal lines.
However, the total lengths occupied by the peaks of the $\phi(x)$ and $\Phi(x)$
fields are different at a given level. The length parameterization in 1D 
means that the comparison of the numbers of peaks  must be done 
at the same total length of the excursion sets 
which generally happens at different levels for the 
$\phi(x)$ and $\Phi(x)$ fields.
For instant, the length of the excursion set of the Gaussian 
field at $\phi(x)=1.95$ (the sum of two heavy line segments at level $=1.95$) 
equals the length of the excursion set
of the non-Gaussian field at $\Phi(x)=1$ 
(the sum of four heavy line segments at level $=1$) . 
One can see at this total
length the number of peaks in the Gaussian field
is two while in the non-Gaussian field it is four. Thus,
the number of peaks parameterized by the total length does
detect this type of non-Gaussianity while the number of
peaks parameterized by the level does not.  Similarly, 
the area 
parameterization of the genus in two dimensions does
detect this type of non-Gaussianity while the level
parameterization of the genus statistic does not.

If the mapping (eq.~\ref{mapping}) is area-preserving than the genus 
of the $\Phi(X,Y)$ field remains exactly Gaussian also in 
the area parameterization. Thus, a detection of such non-Gaussianity
requires additional tests.

We use two different parameterization of the measured morphological 
characteristics and show that the parameterization by the total area of the
excursion set (\ie the cumulative probability function)
has an advantage over parameterization by the temperature
because it gives considerably smaller correlations between different
measures. 
In particular, the total area parameterization detects non-Gaussianity by the
genus statistic in the above example while the temperature parameterization
does not.
For additional discussion of this issue, see \citet{sh02}.

Along with the \qmask map (Xu \etal 2001), we analyze a thousand reference maps
with the same sky coverage, noise properties and power spectrum 
as the \qmask map. For each map
we compute seven morphological functions at a hundred temperature/area levels.
These functions are: 
\begin{enumerate}
\item the total area of the excursion set $A(\dt)$, 
\item the total contour length $C(\dt)$ and $C(A)$ , 
\item the genus of the excursion set $G(\dt)$ and $G(A)$ , 
\item the area of the largest (by area) region $A_p(\dt)$ and $A_p(A)$ ,
\item its perimeter $C_p(\dt)$ and $C_p(A)$, 
\item genus $G_p(\dt)$ and $G_p(A)$, and finally 
\item the number of regions $N_c(\dt)$ and $N_c(A)$ .
\end{enumerate}
The first three are the global \mf and  the following three are the
\mf of the largest (by area) region. 
The largest region at some threshold spans the whole map, or 
``percolates''. 
Its \mf are the best indicators of the percolation phase
transition \citep{sh83,yes-sh96}. Our notations reflect some of
the jargon used in cluster analysis: $A_p, C_p, G_p$ for the \mf
of the percolating region and $N_c$ for the number of regions,
often referred to as ``clusters'' in cluster analysis.

We then calculate the mean for each of these fourteen functions
($\bar{M_i}(\dt)$ and $\bar{M_i}(A),~ i=1,\cdots,7$)  computed
from a thousand Gaussian maps and quantify the differences of every 
function for every map with respect to the mean functions in 
quadratic measure. Finally,
we compute these differences for the \qmask map and test
the Gaussianity hypothesis by comparing them to
the thousand reference maps. 

The rest of the paper is organized as follows. In Sec. 2 and 3 we briefly
describe the \qmask map and the Gaussian simulations thereof. In Sec.~4
we define the morphological functions, their parameterization and
the method of quantifying the departure from the Gaussianity. In Sec.~5
we summarize the results. 

\section{\qmask Map} 

\newcommand{\x}{{\bf x}}
\newcommand{\z}{{\bf z}}
\newcommand{\zero}{{\bf 0}}
\newcommand{\rh}{\widehat{\bf r}}
\newcommand{\C}{{\bf C}}
\newcommand{\bfL}{{\bf L}}
\newcommand{\I}{{\bf I}}
\newcommand{\N}{{\bf N}}
\newcommand{\bfS}{{\bf S}}
\newcommand{\W}{{\bf W}}
\newcommand{\el}{\ell}
\def\expec#1{\langle#1\rangle}

The \qmask map, described in \citep{xu-etal01}, 
combines all the information from the QMAP
\citep{qmap1,qmap2,qmap3} and Saskatoon \citep{Netterfield95,Netterfield97,saskmap}
experiments into a single map at 30-40 GHz covering about 
648 square degrees around the North Celestial Pole.
The map consists of sky temperatures in 6495 sky pixels,
conveniently grouped into a 6495-dimensional vector $\x$,
with a FWHM angular resolution of $0.68^\circ$.
As detailed in \citet{xu-etal01}, all the complications of the
map making and deconvolution process are encoded in the corresponding 
$6495\times 6495$ noise covariance matrix $\N$.
The map has a vanishing expectation value $\expec{\x}=\zero$ and a covariance matrix
given by 
\begin{equation}
\C\equiv \expec{\x\x^t} = \N + \bfS,
\end{equation}
where the signal covariance matrix $\bfS$ is given by
\begin{equation}
\bfS_{ij} = \sum_{\el=2}^\infty {\el(\el+1)\over 4\pi} 
P_\el(\rh_i\cdot\rh_j) e^{-\theta^2\el(\el+1)} C_\el, 
\end{equation}
the $P_\el$ are Legendre polynomials, $C_\el$ is the angular power spectrum,
$\rh_i$ is a unit vector pointing towards
the $i^{\rm th}$ pixel, 
and $\theta=$FWHM$/\sqrt{8\ln 2}$ is the rms beam width.
When computing $\bfS$ in practice, we use the smooth 
power spectrum of \citep{qmaskpow} that fits the observed 
QMASK power spectrum measurements.

Since the raw map has a large and complicated noise component,
we work with the Wiener-filtered version of the map in this paper, 
shown in Fig. \ref{QMASK_map} and defined as
\begin{equation}
\x_w\equiv \W\x, \quad \W\equiv \bfS(\N + \bfS)^{-1}.
\end{equation}
This approach was also taken in Park et al (2001) and Wu et al (2001).
 
\section{Mock Maps}

In order to quantify the statistical properties of our Minkowski functionals,
we need large numbers of simulated QMASK maps.
We therefore generate one thousand mock Gaussian
maps as follows. 
The covariance matrix $\C_w$ of the Wiener-filtered map is given by
\begin{equation}
\C_w\equiv \expec{\x_w x_w^t} = \expec{(\W\x)(\W\x)^t} = \W\expec{\x\x^t}\W^t
= \W\C\W^t,
\end{equation}
and we Cholesky-decompose it as $\C_w=\bfL\bfL^t$, where 
the matrix $\bfL$ lower triangular.
It is straightforward to show that if $\z$ is a vector of independent
Gaussian random variables with zero mean and unit variance (which is
trivial to generate numerically), {\it i.e.},
$\expec{\z\z^t}=\I$, the identity matrix, then 
the mock maps defined $\x_m\equiv\bfL\z$ will have a multivariate Gaussian
distribution with $\expec{\x_m\x_m^t} = \C_w$.

\section{Morphological Statistics}

In this paper we use a set of morphological statistics based on
Minkowski functionals. As suggested by \citep{nov-fel-sh99}, one can
use both global and partial Minkowski functionals in studies of 
CMB maps. In this study we use the global \mf and those of the 
largest region by area (the so-called percolating region).
   
\subsection{Global \mf}
Global Minkowski functionals were introduced into cosmology as
quantitative measures of CMB anisotropy by \citet{got-etal90}
although without reference to the \mf and thus without stressing
their unique role in differential and integral geometry.
\citet{mec-etal94} were the first to place the studies of morphology
of the large scale structure in the context of differential and
integral geometry. In particular, they emphasized a powerful theorem 
by \citep{had57}
stating that with rather broad restrictions, the set of \mf 
provides a complete description of the morphology 
\citep[for further discussion see][]{ker99}. 

Global Minkowski functionals describe the morphology of the excursion set 
at a chosen threshold: $A(\dt)$, $C(\dt)$ and $G(\dt)$ 
are the total area, contour  length and
genus of the excursion set, respectively. The \mf  of Gaussian fields
are known analytically as functions of the threshold.
Assuming that the field is normalized, \ie $u=(\dt-<\dt>)/\sigma_{\dt}$ 
so that $<u>=0, <u^2>=1$, the \mf are
\begin{eqnarray}
A(u)&=&\frac{1}{2} \left[1-{\rm erf}\left(\frac{u}{\sqrt{2}}\right)\right],\cr
C(u)&=&\frac{1}{2R}\exp\left(-\frac{u^2}{2}\right),\label{GMF_G}\cr
G(u)&=&\frac{1}{(2\pi)^\frac{3}{2}} \frac{1}{R^2}u 
\exp\left(-\frac{u^2}{2}\right),
\end{eqnarray} 
where  $R= \sqrt{2}/\sigma_1$
is the characteristic scale of fluctuations in the field;
$\sigma_1$ is the rms of the first derivatives (in
statistically isotropic fields both derivatives $\partial u/\partial x$
and $\partial u/\partial y$ have equal rms). 
$A(u)$ is the fraction of
the area in the excursion set and thus equals the cumulative
probability function: $A(u)=\int_u^{\infty}p(u')du'$, where $p(u)$ is
the probability function. $C(u)$ and 
$G(u)$ are the length of the contour and the genus of the excursion
set per unit area respectively. 

In this study we use somewhat different normalization and units.
We measure the total contour length in the mesh units
and we define $G$ as the number of isolated regions minus the
number of holes in the excursion set.

\subsection{Percolating region}
The morphology of every isolated region in the excursion set can
be characterized by three partial \mf: the area, boundary contour length
and genus of the region. 
In order to describe the morphology of the field in more detail,
\citet{nov-fel-sh99} used the distribution functions of the partial
\mf at several level thresholds.  \citet{bha-etal00} used the averaged
shape parameters in the study of the morphology in the LCRS slices. 
Here we choose to utilize only the \mf of the largest
region ($A_p(\dt), C_p(\dt), G_p(\dt)$)  in addition to the global \mf. 
This is dictated by the relatively
small size of the \qmask map: the maximum number of isolated regions is
about 30 and therefore the distribution functions are very noisy. 
The \mf of the largest, \ie, percolating region of Gaussian fields
are not known in an analytic form but can be easily measured in 
Monte Carlo simulations \citep{sh02}.

\begin{figure}
\includegraphics[height=15cm]{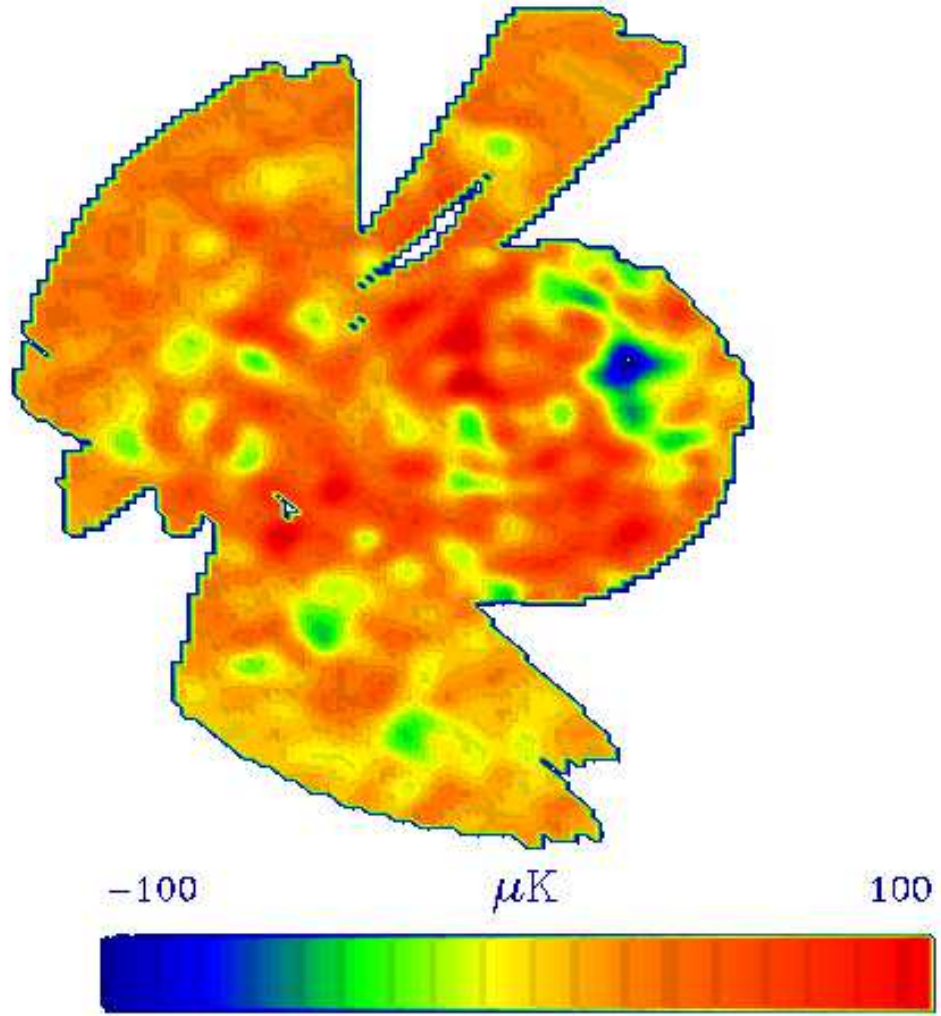}
\caption{Grey scale \qmask map. Light color correspond to higher temperatures.
Note that there are three clearly distinct white regions in the map 
where data are absent.} \label{QMASK_map}
\end{figure}

An additional comment may be useful. A naive idea
of percolation invokes an image of the region connecting the oposite sides of
the map. Thus, the question may arise: what does percolation mean in a 
map with complex boundaries and holes like the \qmask map 
(Fig. \ref{QMASK_map}).
In particular, which points must be connected at the percolation
threshold in a map with irregular boundaries? 
The fact is that percolation is a phase
transition which takes place regardless of the shape and topology of the map
(see \eg \citet{sta-aha92}).
It can be studied and relatively easily measured in maps of arbitrary shapes
including holes. 
The \mf of the largest region are particularly useful while the direct
detection of the connection between the oposite sides is much more
unstable even in simple square maps \citep{dom-sh92}. 
At percolation
transition the area and perimeter of the largest region 
experience a rapid growth while the genus rapidly decreases
\citep{sh02}. 
Similarly to many other statistics the percolation
transition is affected mostly by the size of the map 
and its effective dimensions and much less by its
shape or presence of holes. For example, if a three-dimensional map has 
the shape of a thin wedge  then percolation becomes two-dimensional
\citep{sh-yes98}. Likewise, a two-dimensional map having the shape of 
a narrow strip effectively percolates as a one-dimensional map.
More accurately, if the smallest size of the map is smaller than the
scale of the field then the percolation transition effectively reduces
the dimensions. A simple visual inspection of Fig. \ref{QMASK_map} shows that nothing
of the kind of the problems discussed above is present in the \qmask map.
The major drawback of the map is its size not the shape.

In addition, we wish to stress that we do not compare
the percolation curves measured for the \qmask map with the standard
Gaussian curves. Instead we compare them with the curves measured 
in the reference
maps having exactly same irregular shape and holes as the \qmask map.
Therefore, the question of accuracy of reproduction of the standard percolation
properties in two-dimensional maps becomes largely irrelevant.

\subsection{Numerical technique}
The numerical technique used for measuring the \mf is described in detail
in \citep{sh02}. Here we outline the main features that may differ from
numerical methods used by others \citep{col88,got-etal90,sch-buc97,win-kos97,sch-gor98,par-etal01,wu-etal01}. 

First, we use neither
Koendernik invariants \citep{koe84} nor Croftons's formulae 
\citep{cro68} which are often used in other studies.
A pixelized map is considered to be an approximation to a continuous field. 
This approach was used by \citet{nov-fel-sh99} but here we use a different and
more refined version of the code computing \mf. The major effect of this
improvement is the numerical efficiency of the code.

\begin{figure}
\includegraphics[height=15cm]{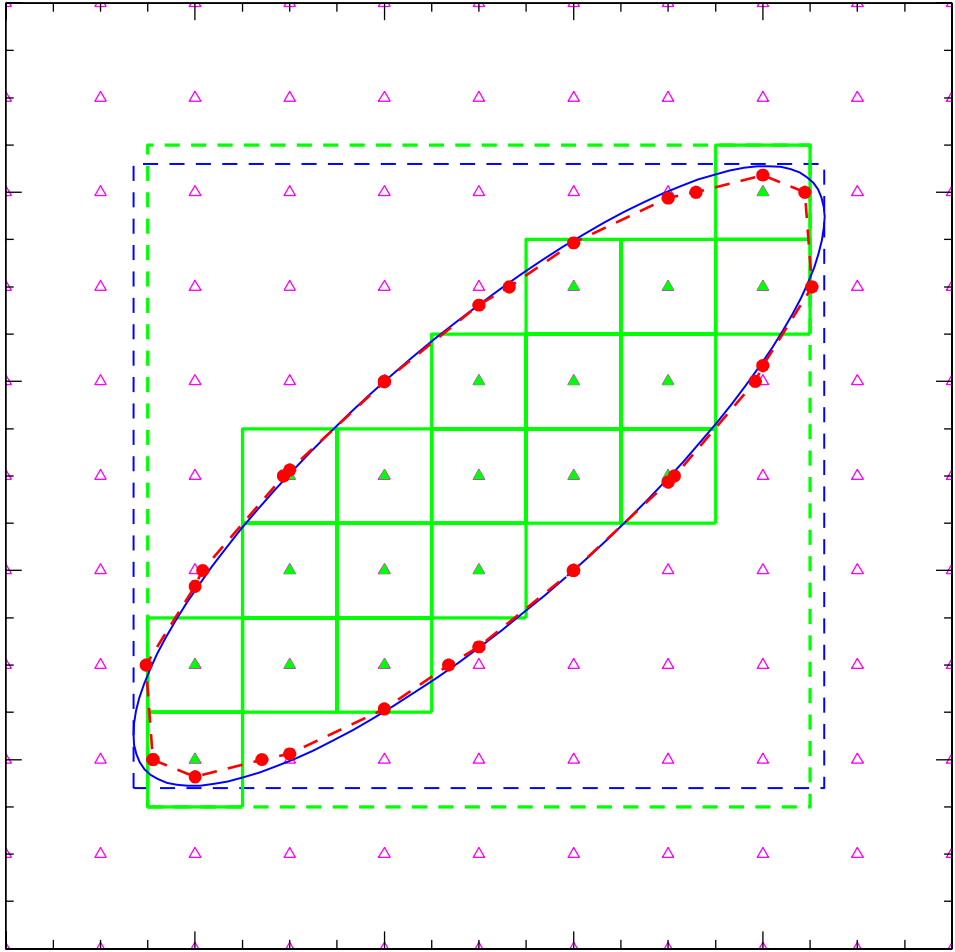}
\caption{An illustration of the numerical technique.
The solid line shows the elliptical contour corresponding to the certain
threshold $u=u_{th}$. Triangles mark the sites of the lattice: 
solid with $u > u_{th}$ and empty with $u < u_{th}$. 
Solid circles  mark the contour points satisfying the condition
$u = u_{th}$ (the roots of eq. \ref{inter}); the dashed line
is the resulting contour.
The true area of the region is the area within the solid ellipse.
We approximate it by the area within the dashed contour while in most 
works it is approximated by the sum of areas of the elementary squares
(solid squares).
We approximate the perimeter by the length of the dashed contour
while in other works it is often approximated by the sum of the
external edges of the elementary squares. One can easily see that 
this approximation gives the value of 
the perimeter of the large dotted rectangle. As the lattice constant
approaches zero it converges to the perimeter of the large solid rectangle,
while our perimeter converges to the true value of the perimeter of
the solid ellipse.}  \label{NumMeth}
\end{figure}

The major features of our technique are illustrated by Fig. \ref{NumMeth}. 
Suppose we use a regular square lattice and the true contour
corresponding to $u=u_{th}$ is a simple ellipse shown by the solid line
in Fig. \ref{NumMeth}. The sites satisfying the threshold condition $u>u_{th}$
are shown by filled triangles, while the sites with $u<u_{th}$
are shown by empty triangles. The union of elementary squares (shown by
heavy solid lines) placed on each site satisfying the
threshold condition is widely used as an approximation to the region. 
One can easily count the number of
elementary squares, number of edges of the elementary squares
and number of vertices in the union.
Then by using Crofton's formulae one computes the area, perimeter and 
the Euler characteristic of the region. For instance, the area in this 
approximation is simply the sum of areas of the elementary squares. 
The perimeter is the
number of external edges of the elementary squares. While the area 
converges to the right value as the lattice constant
approaches zero, the perimeter does not (for a detailed discussion
of this effect see \eg \citet{win-kos97}). The perimeter of the 
set of the elementary squares obviously equals the perimeter of the
rectangular region shown by the dotted line. As the lattice parameter
approaches zero the perimeter of the region approches the perimeter
of the large rectangle shown by the solid line which obviously 
is a wrong value. 
Assuming the isotropy of the
map one can show that the mean length of a segment is $4/\pi$ times
larger than its true value \citep{win-kos97} and can be corrected
only statistically. Our method is completely free of this drawback and 
does not need such a correction.     

In our approach we begin with the construction of  the contour points
for each isolated region of the excursion set.
Suppose that $u(i,k) < u_{th}$ but  $u(i+1,k) \ge u_{th}$ then
the contour point lies somewhere between ($i,k$) and ($i+1,k$) sites.
Its coordinates ($x,y_k$) can be obtained by solving the 
linear interpolation equation 
\begin{equation}
u_{th} = u(i,k) +\frac{u(i+1,k)-u(i,k)}{x(i+1,k)-x(i,k)} 
(x-x(i,k)) \label{inter}
\end{equation}
for the coordinate $x$. 
We construct the contour points by linear interpolation of the field between
the sites in both horizontal and vertical directions. These points are 
marked by the solid circles in Fig. \ref{NumMeth}. 
Then, computing the perimeter 
we assume that the contour points are connected by the segments of 
a straight line (dashed line).  
The contours made by this method converges to the right value
of the perimeter as the lattice constant approaches zero.
 
\begin{figure}
\includegraphics[height=15cm]{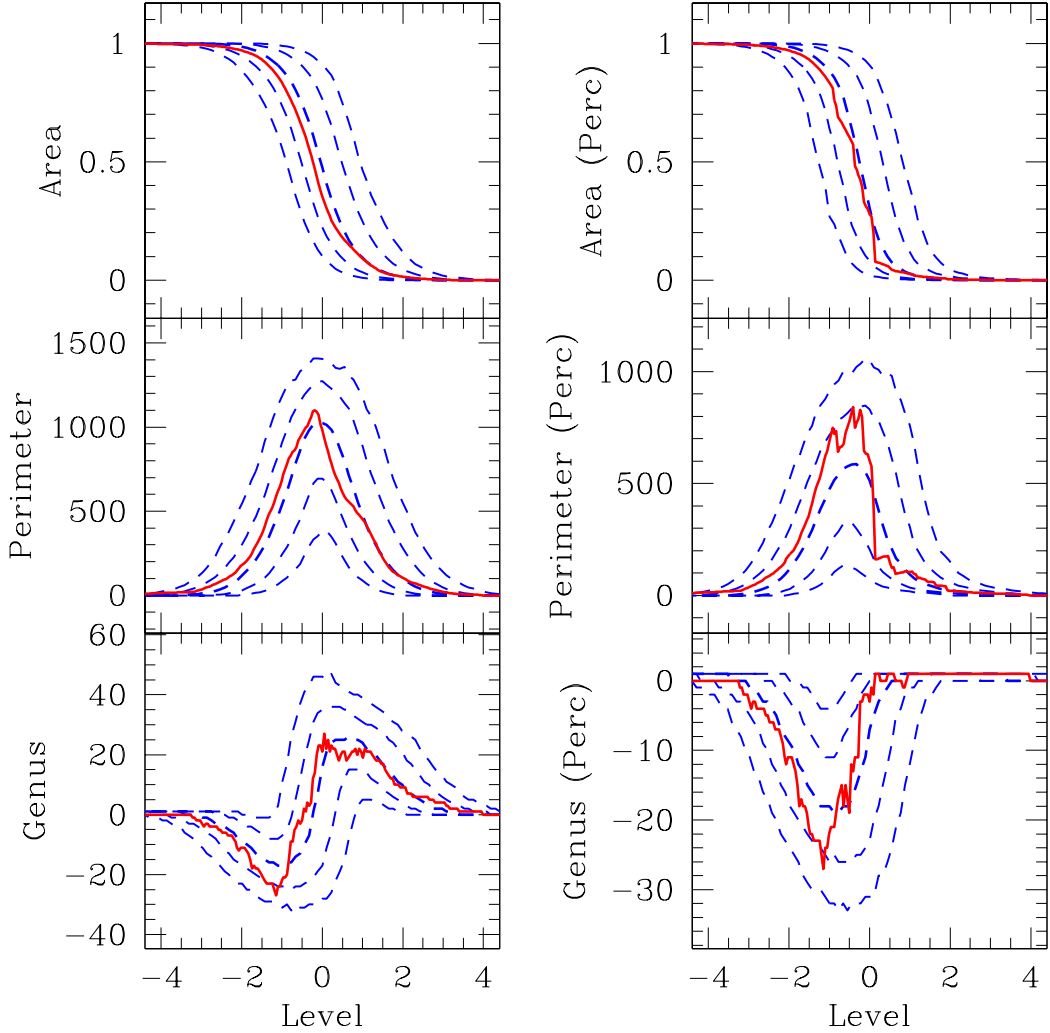}
\caption{\mf as functions of $u=\dt/\sigma_{\dt}$ for the excursion set
(left hand side column) and for the percolating region 
(right hand side column). 
Top row: the area fraction $A$ and $A_p$, 
middle row : contour length $C$ and $C_p$,
bottom row: genus $G$ and $G_p$.
The contour lengths are given in mesh units. The genus is
``number of regions'' $-$ ``number of holes''. 
The solid lines show the parameters of the \qmask map,
heavy dashed lines show the median Gaussian values, thin dashed lines
show 68\% and 95\% ranges.}  
\label{LevelParam}
\end{figure}

The other major difference consists in allowing a direct linkage
of diagonal grid sites in some cases depending on the values in the four
neighboring sites. 
Consder a case when two diagonally adjacent pixels satisfy the
threshold condition: $u(i,k) \ge u_{th}$ and $u(i+1,k+1) \ge u_{th}$
but the sites $u(i+1,k)<u_{th}$ and $u(i,k+1)<u_{th}$. 
One usually chooses either consider the 
$(i,k)$ and $(i+1,k+1)$ sites always directly linked or always not linked 
implying that the difference 
between the two alternatives disappears with vanishing the lattice 
parameter.
Our definition of the direct linkage of two diagonally adjacent pixels 
is different from both. The diagonally adjacent pixels are directly 
linked if
$u_{mean}=(u(i,k)+u(i+1,k)+u(i,k+1)+u(i+1,k+1))/4 \ge u_{th}$
and directly not linked otherwise. In the latter case they of course 
are allowed  to be linked by the friend of friend procedure if they 
have proper friends. 
This method affects the number of regions, their sizes and the number
of holes in the regions and therefore the values of all \mf
(including genus). The visual inspection of Fig. \ref{LevelParam} and
Fig. 4 in \cite{par-etal01} reveals small but noticeable differences
in the genus curves. This treatment of the direct diagonal linkage also 
considerably reduces the intrinsic anisotropy of a rectangular grid.    
Since the \qmask map occupies a small part of the sky, we ignore
the intrinsic curvature of the map when computing the areas and perimeters.

If a part of the region boundary is the boundary of the map mask, then
it does not contribute to the contour length of the region.
Similarly, the holes in the map mask contribute neither into the genus
nor the contour length.

The analysis of one realization of a
Gaussian random field at a hundred levels takes about 0.7s, 2.9s,
12.3s, and 73s for $64^2$, $128^2$, $256^2$, and $512^2$ maps
respectively on HP C240 workstation. It corresponds approximately to
the $t \propto N_{pix}\ln(N_{pix})$ dependence of the computational time on
the size of the map.

\subsection{Parameterization by the level of $\dt$.}
In addition to six \mf, we computed the number of isolated regions $N_c$
(Novikov, Feldman, \& Shandarin 1999).
Thus, in total we compute seven morphological parameters 
\begin{equation}
M_i(\dt)=[A(\dt),C(\dt),G(\dt),A_p(\dt),C_p(\dt),G_p(\dt),N_c(\dt)], ~~~
i=1,...,7  \label{mfs_level}
\end{equation}
at a hundred threshold levels $\dt$ for the \qmask map and the
thousand reference maps.
The six panels of Fig. \ref{LevelParam} show six \mf measured for the 
\qmask map (solid lines) and the mean values for a thousand Gaussian
maps (heavy dashed lines). The thin dashed lines show the 68\% and 
95\% intervals. 

The genus curve (left bottom panel) is in good qualitative agreement with
that obtained by \citet{par-etal01}. However, it is worth noting that we use 
a different normalization and a somewhat different variable on the 
horizontal axis:
we use the level normalized to the rms $u=\dt/\sigma_{\dt}$, 
while  Park \etal use a more
complex quantity $\nu_A$ related to the excursion set of the Gaussian field.
Park \etal describe their parameterization as follows:
``the area fraction threshold level $\nu_A$ is defined to be the 
temperature threshold level at which the corresponding isotemperature contour
encloses a fraction of the survey area equal to that at the temperature level 
$\nu_A$ for a Gaussian field'' (the bottom of p.586 ). Some quantitative 
differences may be also ascribed to a different linkage scheme used
by Park \etal (see the previous section). We wish to stress that the 
parameterization by
$\nu_A$ has similar properties (including the noise correlation properties)
with the parameterization by the area ($A$)
discussed in the following section. The only difference between them is the 
geometrical interpretaion. 

\begin{figure}
\includegraphics[width=12cm]{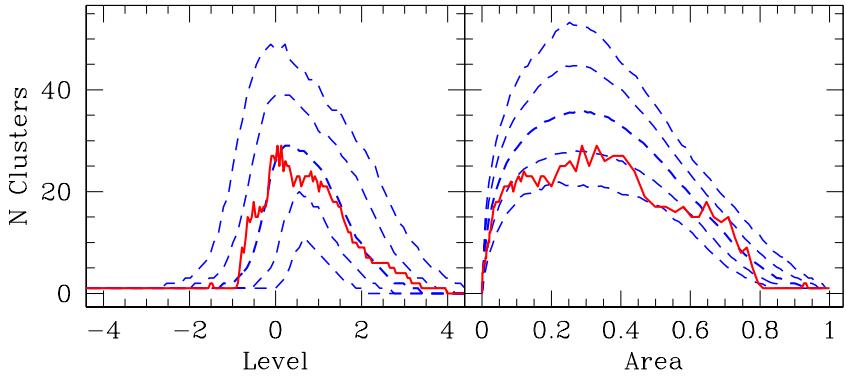}
\caption{The number of regions $N_c$ as a function of the level
$u=\dt/\sigma_{\dt}$ (left hand
panel) and as a function of $A$ (right hand panel).
Other notation is as in Fig. \ref{LevelParam}}  \label{Ncl}
\end{figure}

The number of isolated
regions is shown in Fig. \ref{Ncl} for both the temperature and global area
parameterizations. The latter is described in detail in Sec. 4.5.

In order to quantify the degree of non-Gaussianity of the \qmask map,
we define and compute a functional that quantifies the ``distance'' 
between two functions in the functional space of measured statistics. 
The distance is defined by the integral
\begin{eqnarray}
D_{ik}^{(\dt)}&=&\left\{ \int_{-\infty}^{\infty}\left[M_{ik}(\dt)-\bar{M_i}(\dt)\right] ^2~d(\dt)\right\}^{1/2},\cr
i&=&1,...,7; ~~~k=1,...,1000 \label{dis_lev}
\end{eqnarray}
where 
\begin{equation}
\bar{M_i}(\dt)= \frac{1}{1000}\sum_{k=1}^{1000} M_{ik}(\dt) 
\end{equation}
is the mean value of the statistic as a function of $\dt$.
The integral in eq.~(\ref{dis_lev}) is approximated by the sum over 
a hundred levels of $\dt$. We also compute the distances of the each
morphological curve from the Gaussian mean for the \qmask map.
The third column of Table 1 shows the percentile of the \qmask 
map distances from the mean for each morphological statistic
parameterized by $\dt$. The highest departure
from the mean is shown by the $A(\dt)$ curve which is the cumulative
probability function. It shows that only about 27\% of Gaussian
curves are closer to  the Gaussian mean than the \qmask curve. 
Other parameters are even closer to the corresponding Gaussian mean.

\subsection{Parameterization by the total area $A$.}
The parameterization of the morphological
parameters by the level of the field $u$ has serious drawbacks
mentioned in Introduction. Although the parameterization by $\nu_A$ used
in most studies of the genus is free of these drwbacks
it seems to be more natural for morphological studies to use
the total area $A(\dt)=A_{ES}(\dt)/A_{m}$ (where $A_{ES}(\dt)$ is the 
area of the excursion set at $\dt$ and $A_{m}\equiv A_{ES}(-\infty)$ 
is the total area of
the map) as an independent parameter \citep{yes-sh96}. 
It is one of the \mf measured for the map, 
it represents the cumulative distribution functions of the field,
and in addition it has a very simple geometrical interpretaaiton.
It also has an additional advantage of decorrelating different
statistics as discussed bellow. The $\nu_A$ parameterization
must have similar properties because it is related to $A$
by a monotonic transform (the first eq. of \ref{GMF_G}). 

\begin{figure}
\includegraphics[width=15cm]{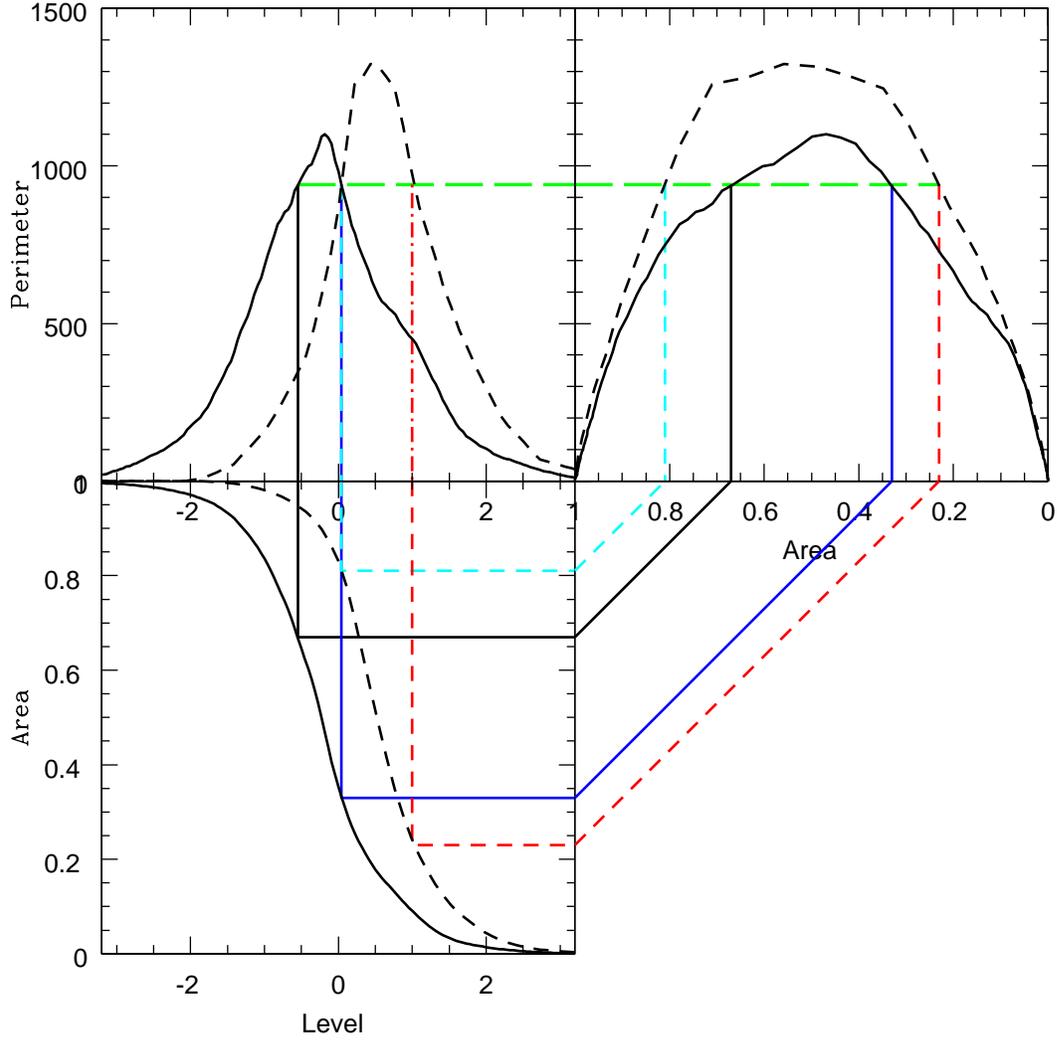}
\caption{Illustration of the transformation from $\dt$ to $A$
parameterization. The thick solid lines show $C=C(\dt)$ (top left panel)
$C=C(A)$  (top right panel) and $A=A(\dt)$ (bottom left panel)
for the \qmask map.
The thin solid lines illustrate the transformation. The dashed lines
show similar transformation for a randomly chosen Gaussian map.}
\label{fLeveltoA}
\end{figure}

The transformation from the level parameterization to the area
parameterization is a highly nonlinear procedure
illustrated in Fig. \ref{fLeveltoA}. 
The thick lines correspond to the \qmask map (solid)
and a randomly chosen Gaussian map (dashed). The transformation from $C(\dt)$
to $C(A)$ involves the $A(\dt)$ function shown in the lower panel,
which is different for each map.
The thin solid lines show the transformation for the \qmask map and
dashed lines for the Gaussian map. Note that in order to illustrate
this transformation, we reversed the $A$ axis in the  $C=C(A)$ plot.

As an illustration of the diference between the level and area
parameterization consider a simple example of a strong nonlinear
field. Let the field $w(x,y)$ is the exponent of  
a Gaussian field $u(x,y)$: $w=\exp(\alpha u)$.
The global \mf of such a field can be easily derived from eq. \ref{GMF_G}
\begin{eqnarray}
A(w) &=&\frac{1}{2} \left[1-{\rm erf}\left(\frac{\ln w}{\alpha \sqrt{2}}\right)\right],\cr
C(w)&=&\frac{1}{2R}\exp\left(-\frac{(\ln w)^2}{2}\right),\label{GMF_LN}\cr
G(w)&=&\frac{1}{(2\pi)^\frac{3}{2}} \frac{1}{R^2} (\ln w)
\exp\left(-\frac{(\ln w)^2}{2}\right). 
\end{eqnarray}

\begin{figure}
\includegraphics[height=15cm]{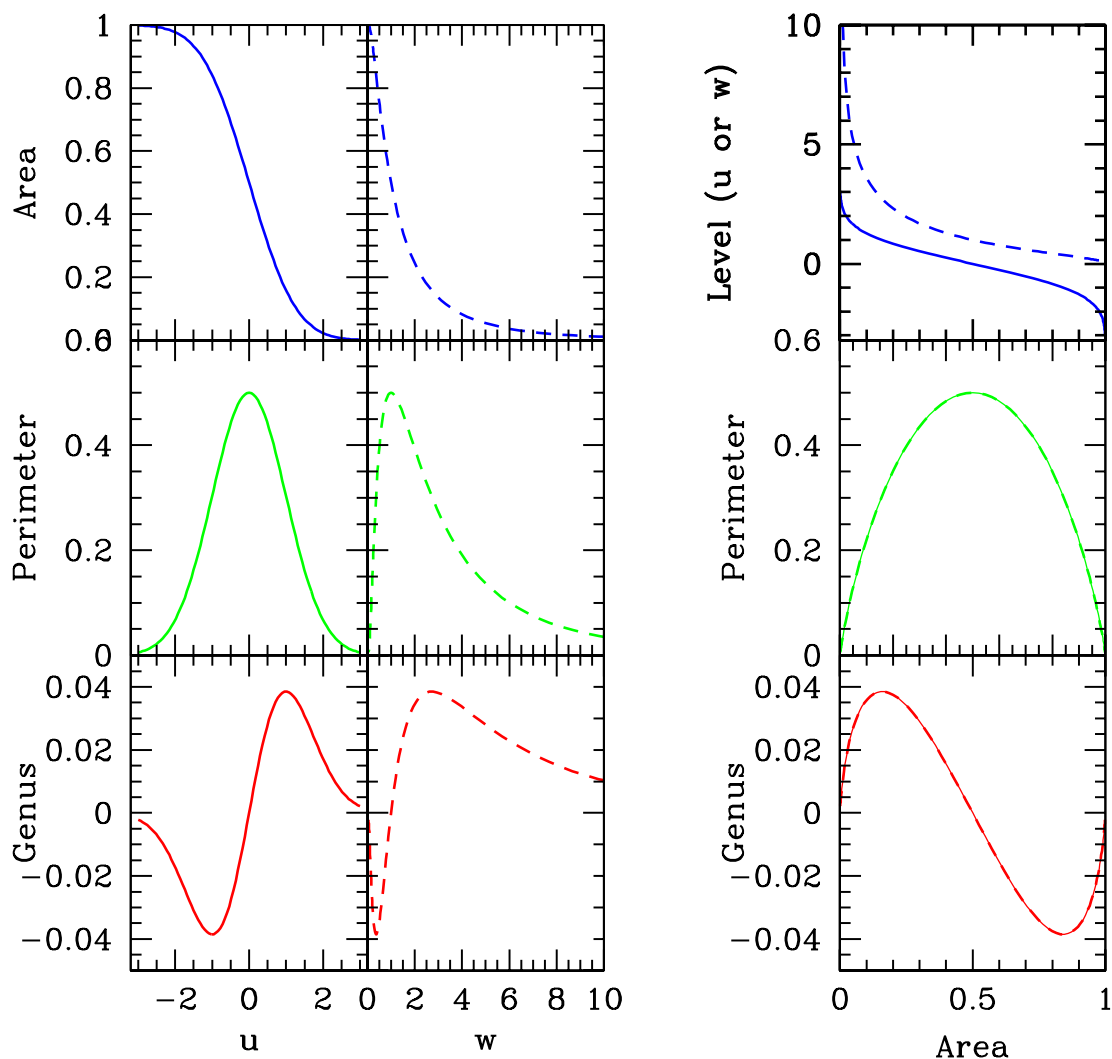}
\caption{The global \mf of the parent  Gaussian field $u$ 
and derived non-Gaussian field $w=\exp u$ are shown in two left hand 
side panels as functions of the level. 
Both the perimeter and genus remain the same for both
fields if they are parameterized by the total area (two bottom
right hand side panels).
All information about the non-Gaussianity of the $w$-field is stored in
the cumulative probability function (dashed lines in the 
top panels). 
The solid lines in the top panels show the Gaussian cumulative 
probability function.} 
\label{GnGParam}
\end{figure}

Fig. \ref{GnGParam} illustrates the diference between  the \mf of the Gaussian 
field $u$ and non-Gaussian field $w$ (with $\alpha=1$) if both are 
parameterized by the level. The percolation curves differ strongly
as well. 
On the other hand both the perimeter and genus as the functions
of the area remain identical for the both fields because the transformation
simply relabels the levels but does not affect the geometry of the contours.
This is an example of a trivial non-Gaussian field when all non-Gaussian 
information is stored in the one-point probability density function.
The further discussion of such fields can be found in \citet{sh02}.

\begin{figure}
\includegraphics[width=15cm]{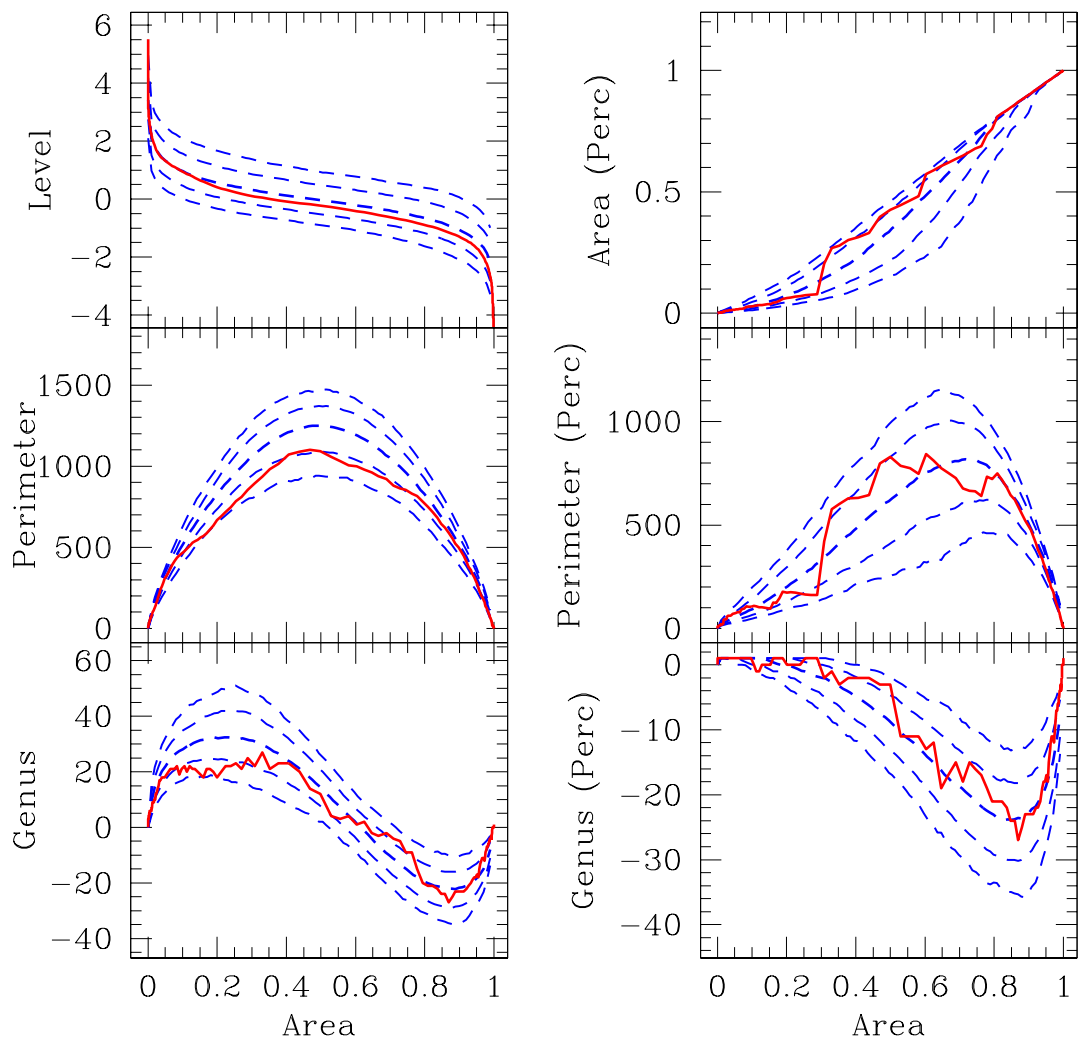}
\caption{The figure is similar to Fig. \ref{LevelParam}, 
except that all morphological
parameters are functions of $A=A_{ES}/A_m$, where $A_{ES}$ and 
$A_m=A_{ES}(-\infty)$ is the area of the excursion set 
and that of the whole map, respectively.} \label{AParam}
\end{figure}

Using the area parameterization, we compute seven new morphological
functions for each map:
\begin{equation}
M_i(A)=[\dt(A),C(A),G(A),A_p(A),C_p(A),G_p(A),N_c(A)],~~~
i=1,...,7. \label{mfs_a}
\end{equation} 
Figure \ref{AParam} shows the measured quantities as functions of $A$.
Similarly to Fig. \ref{LevelParam} six panels show six \mf measured for the 
\qmask map (solid lines) and the mean values for a thousand Gaussian
maps (heavy dashed lines). The thin dashed lines show the 68\% and 
95\% intervals. 

The difference of the \qmask curves from the Gaussian mean  
is quantified similarly to that described before by the distance
between curves in the functional space: 
\begin{eqnarray}
D_{ik}^{(A)}&=&\left\{\int_0^1\left[M_{ik}(A)-\bar{M_i}(A)\right] ^2~dA  
\right\}^{1/2}\cr
i&=&1,...,7; ~~~k=1,...,1000  \label{dis_a}
\end{eqnarray}
where
\begin{equation}
\bar{M_i}(A)= \frac{1}{1000}\sum_{k=1}^{1000} M_{ik}(A).
\end{equation}
There is an important difference between the two distances defined by 
eq. \ref{dis_lev} and \ref{dis_a}. Using $dA=-p_k(\dt)d(\dt)$ where
$p_k(\dt)$ is the estimate of the probability function 
derived from the $k$-th map one can rewrite eq. \ref{dis_a}
as an integral over $\dt$:
\begin{eqnarray}
D_{ik}^{(A)}&=&\left\{\int_0^1\left[M_{ik}(A)-\bar{M_i}(A)\right] ^2~dA \right\}^{1/2}\cr
&=&\left\{\int_{-\infty}^{\infty}\left[M_{ik}(\dt)-\bar{M_i}(\dt)\right]^2 
p_k(\dt)~d(\dt)\right\}^{1/2} \label{dis_comp}
\end{eqnarray}
Comparing eq.~(\ref{dis_lev}) and (\ref{dis_comp}), one notices that
the distance between the curves computed in the area space 
(eq.~\ref{dis_a}) depends 
more on the bulk of the probability function and less on the tails 
of the probability function, since it is weighted by the 
probability function. This obviously makes the measure more robust 
because the measurements at extreme levels are subject
to large fluctuations. 
It is worth stressing that the calculation of all morphological
parameters $M_{ik}$ themselves (eq.~\ref{mfs_level} 
and \ref{mfs_a}) is independent of the type of parameterization
because $M_{ik}(A) \equiv M_{ik}(\dt) $ provided that $A$ is measured
at corresponding $\dt$: $A=A(\dt)$. The only quantities that depend on the
parameterization are the distances from the Gaussian expectation functions 
(eqs.~\ref{dis_lev} and 
\ref{dis_a}) that are used for ranking the maps.

The second column of Table 1 shows the percentile of the 
\qmask distances from
the mean in the area parameterization. Comparing the values 
corresponding to different
parameterizations, one can easily see that the parameterization
by $\dt$ suggests that the \qmask map is a more typical example
of a Gaussian field than that using $A$ as an independent
parameter. The parameterization by the level suggests that the most
exceptional deviation of the \qmask map from Gaussianity is given by the 
$A(\dt)$ curve, but more than 70\% of Gaussian maps differ from 
the mean more than the \qmask map. Thus, the \qmask map looks like a very
typical realization of the Gaussian field.
Parameterizing by the total area, one may conclude
that the Gaussian $P(A)$ and $N_c(A)$ curves
differ more than the \qmask map in only about 21\% 
and 18\% of the cases respectively. 
Thus, this parameterization
reveals that the \qmask map is not quite a typical realization of the
Gaussian field, but is not very unusual either, quite consistent
with Gaussianity. 

It is perhaps worth noting that all statistics but the genus of the 
percolating region, $G_p$ show a greater difference of the \qmask map with
respect to the corresponding expectation value of the set of Gaussian maps
in the $A$ parameterization than that in $\dt$ parameterization. Since
the difference between two parameterizations for $G_p$ is not large
(12\% and 21\%)
it maybe a statistical fluctuation. One also may note that two
columns of Table 1 appear anti-correlated. As we have discussed before
the quantities shown in two columns of Table 1 have different
sensitivity to the measurements at extreme $\dt$ levels both highest and
lowest. The numbers in the first column corresponding to the $A$ 
parameterization are more robust.

\subsection{Cross-correlations}
   
The parameterization of the morphological statistics by the total area
makes the set of parameters $M_i(A)$  (eq.~\ref{mfs_a})
less correlated than $M_i(\dt)$ (eq.~\ref{mfs_level}).
Table 2 shows the correlation coefficients of the distances
\begin{equation}
r_{ij}=\frac{\sum\limits_{k=1}^{1000}(D_{ik}-\bar{D_i}) 
(D_{jk} - \bar{D_j})}
{\left[\left(\sum\limits_{k=1}^{1000}(D_{ik}-\bar{D_i})^2\right)
\left(\sum\limits_{k=1}^{1000}(D_{jk}-\bar{D_j})^2\right)\right]^{1/2}}
\label{correlations}
\end{equation}
for all 21 pairs of morphological statistics.
The values above the diagonal show the correlations
when all statistics were parameterized by the total area of the excursion set,
$A$, while the values below the diagonal were computed using the 
parameterization by the level, $\dt$.
 
The correlation coefficient shows for each pair of statistics
how well the separation from the mean Gaussian curve in one statistic $M_i$
can be predicted from the other one $M_j$.  
The closer $r_{ij}$ to unity, the less independent are corresponding
statistics. In this respect the $\dt$ and $A$ 
parameterizations are very different.
For instance, the lowest correlation in the class
of the level parameterization is $0.82$ between $G_p(\dt)$ and 
$N_c(\dt)$ distances
and 16 out of 21 cross-correlations are greater than $0.9$.
In the class of the area parameterization, only one correlation coefficient
is greater than $0.82$: that between  $G(A)$ and $N_c(A)$ distances ($0.85$).
Seventeen out of 21 pairs correlate at a level lower than $0.5$
thirteen of which at the level lower than $0.3$. 
Although absence of correlation
does not prove the statistical independence of two sets of numbers,
a correlation coefficient approaching unity indicates 
a strong statistical dependence. 
 
\section{Discussion}
We present the results of a study of Gaussianity in the \qmask map.
In agreement with the first study of topology of the degree-scale 
CMB anisotropy \citep{par-etal01}, which was limited to the genus of QMASK,
we conclude that the \qmask map 
is compatible with the assumption of Gaussian $\dt/T$ fluctuations. 
In this study, we used six morphological statistics 
in addition to the genus, $G$:
the areas $A$ and $A_p$ of the excursion set and the largest (\ie percolating)
region, the contour lengths or perimeters $C$ and $C_p$ of the 
excursion set and the largest region; the genus $G_p$ of the percolating
region and the number of isolated regions $N_c$. 
According to its morphology, the \qmask map  
is not a very typical example of a Gaussian field, but not very exceptional
either: about 20\% of a thousand Gaussian maps have greater differences
with at least one mean function (Table 1). 

We show that the parameterization of the morphological statistic is
very important. A naive parameterization by the temperature threshold
$\dt$ results in a highly correlated (almost degenerate) set of morphological
characteristics: 16 out of 21 cross-correlation coefficients
are greater than 0.9 and all the rest are greater than 0.82
(Table 2, bellow the diagonal).
Representing the same morphological parameters as functions of
the total area of the excursion set $A$ helps to decorrelate them:
now 13 pairs correlate less than 0.3, four more correlate less than 0.5
and the rest less than 0.85 (Table 2, above the diagoanl).

Our measurements of the deviations of the morphological functions from
the Gaussian mean strongly rely on the bulk of the probability function
(see eq. \ref{dis_comp}) 
and therefore are robust to spurious effects.

In the $A$ parameterization the strongest deviations of the 
\qmask map from the
Gaussian expectation values were shown by the number of peaks, $N_c(A)$ 
(82\% \ie the \qmask map deviates from the Gaussian expectation number
of peaks more than 82\% of Gaussian maps),
total contour length of the excursion set, $C(A)$ (79\%) and 
area of the percolating region, $A_p(A)$ (55\%) 
while in the $\dt$ parameterization
by the total area of the excursion set \ie the cumulative probability
function, $A(\dt)$ (27\%), 
genus of the percolating region, $G_p(\dt)$ (21\%)
and contour length of the percolating region, $C_p(\dt)$ (20\%).

The techniques that we have described here are computationally efficient
($\propto O(N_{pix}\ln N_{pix})$),
and should be useful also for much larger upcoming 
datasets such as that from the
MAP satellite. Another promising area for future applications 
is computing Minkowski functionals from foreground maps
and topological defect simulations, thereby enabling
quantitative limits to be placed on the presence of such structures 
in CMB maps.

 
\section{Acknowledgment}
Support for this work was provided by
NSF grant AST00-71213,
NASA grant NAG5-9194,
the University of Pennsylvania Research Foundation, and
the University of Kansas General Research Fund. 
HAF wishes to acknowledge support from the National Science Foundation
under grant number AST-0070702 and the National Center for
Supercomputing Applications. We wish also to thank the referee for a 
constructive criticism and useful suggestions.

\clearpage

\begin{deluxetable}{|l|c|c|}
\tabletypesize{\scriptsize}
\tablecaption{Percentage of Gaussian maps deviating less than QMASK. 
\label{tbl-1}}
\tablewidth{0pt}
\tablehead{
\colhead{} & \colhead{Parameterized by Area, $A$} & \colhead{ Parameterized by Level, $\dt$} 
}
\startdata
Level, $\dt$             & 28.2\% & \\ \hline
Area, $A$                &        & 27.2\% \\ \hline
Perimeter, $C$           & 78.8\% &  6.1\% \\ \hline
Genus, $G$               & 48.3\% &  2.0\% \\ \hline
Area (Perc), $A_p$       & 54.6\% & 12.1\% \\ \hline
Perimeter (Perc), $C_p$  & 24.3\% & 19.7\% \\ \hline 
Genus (Perc), $G_p$      & 11.6\% & 21.4\% \\ \hline
Number of regions, $N_c$ & 82.5\% &  0.4\% \\ 
\enddata
\end{deluxetable}

\clearpage

\begin{deluxetable}{||l||c|c|c|c|c|c|c||}
\tabletypesize{\scriptsize}
\tablecaption{Correlations between different  non-Gaussianity statistics parametrized by temperature, $\dt$ (below the diagonal) and
area, $A$ (above the diagonal).\label{tbl-2}}
\tablewidth{0pt}
\tablehead{
\colhead{} &  \colhead{$A$} & \colhead{$C$}  & \colhead{$G$}  & \colhead{$A_p$}& \colhead{$C_p$}& \colhead{$G_p$}& \colhead{$N_c$} 
}
\startdata
$\dt$&      & 0.08 & 0.15 & 0.06 & 0.10 & 0.13 & 0.16 \\ \hline
$C$  & 0.98 & 1    & 0.66 & 0.13 & 0.23 & 0.34 & 0.49 \\ \hline
$G$  & 0.92 & 0.96 & 1    & 0.30 & 0.22 & 0.44 & 0.85 \\ \hline
$A_p$& 0.97 & 0.97 & 0.93 & 1    & 0.58 & 0.37 & 0.30 \\ \hline
$C_p$& 0.89 & 0.92 & 0.91 & 0.91 & 1    & 0.56 & 0.08 \\ \hline
$G_p$& 0.88 & 0.91 & 0.93 & 0.90 & 0.94 &  1   & 0.12 \\ \hline
$N_c$& 0.92 & 0.94 & 0.96 & 0.92 & 0.84 & 0.82 & 1    \\ 
\enddata
\end{deluxetable}
\clearpage

\end{document}